\documentclass{aa}  
\usepackage{graphicx}
\usepackage{natbib}

\newcommand{\rpup} {$\rho$~Pup}
\newcommand{\dxcet} {DX~Cet}
\newcommand{\kms}{km\,s$^{-1}$}
\newcommand{\Ro}{R_{\odot}}
\bibpunct{(}{)}{;}{a}{}{,}
\usepackage[varg]{txfonts}
\usepackage{color}

\begin{document}

   \title{Understanding the dynamical structure of pulsating stars.}

   \subtitle{HARPS spectroscopy of the $\delta$ Scuti stars $\rho$~Pup
and DX~Cet
\thanks{This work is based on
observations made with the 3.6m telescope at La Silla Observatory under the
ESO Large Programme LP185.D-0056.}
}

   \author{N.~Nardetto\inst{1}\and E.~Poretti\inst{1,2}\and M.~Rainer\inst{2}\and 
           G.~Guiglion\inst{1}\and M.~Scardia\inst{2}\and V.S.~Schmid\inst{3} \and P.~Mathias\inst{4,5} 
          }

   \institute{Laboratoire Lagrange, UMR7293, Universit\'e de Nice Sophia-Antipolis, CNRS, Observatoire de la C\^ote d'Azur, Nice, France  \\
              \email{Nicolas.Nardetto@oca.eu}
         \and
%\institute
 INAF -- Osservatorio Astronomico di Brera,
              Via E. Bianchi 46, 23807 Merate (LC), Italy
\and
 Instituut voor Sterrenkunde, KU Leuven, Celestijnenlaan 200D, B-3001 Leuven, Belgium \and
 Universit\'e de Toulouse, UPS-OMP, IRAP,  F-65000 Tarbes, France
\and
CNRS, IRAP, 57, Avenue d'Azereix, BP 826, F-65008 Tarbes, France 
% Institut de Recherche en Astrophysique et Plan\'etologie, CNRS, 14 avenue Edouard Belin, Universit\'e de Toulouse, UPS-OMP, IRAP, 31400 Toulouse, France
}

   \date{Received April 23, 2013; accepted ...}

% \abstract{}{}{}{}{} 
%  {} token are mandatory
 
  \abstract
   {High-resolution spectroscopy is a powerful tool to study the dynamical structure of pulsating stars atmosphere.  }
   {We aim at comparing the line asymmetry and velocity of the two $\delta$ Sct stars $\rho$~Pup and DX~Cet with previous spectroscopic data obtained on classical Cepheids and $\beta$ Cep stars.}
   {We obtained, analysed and discuss HARPS 
%(High Accuracy Radial velocity Planetary Search) 
high-resolution spectra of  $\rho$~Pup
and DX~Cet. We derived the same physical quantities as used in previous studies, which are the first-moment radial velocities and the bi-Gaussian spectral line asymmetries.  }
   {
The identification of $f$=7.098~d$^{-1}$ as a fundamental radial mode and the very accurate {{\it Hipparcos}} 
parallax promote \rpup\, as the best standard candle to test the period-luminosity relations of
$\delta$~Sct stars. The action of small-amplitude nonradial modes can be seen as well-defined cycle-to-cycle 
variations in the radial velocity measurements of \rpup.
%We definitively conclude that cycle-to-cycle variations in the {\bf radial velocity measurements of \rpup\, are 
%not related to chromospheric activity, but to small-amplitude nonradial modes. 
%due to to small-amplitude nonradial modes.} 
%\ep{[REALLY NECESSARY?] They affect the determination of the amplitude of the radial velocity curve of the fundamental  mode up to 3\%
%to the large amplitude of the radial mode, the nonradial modes have a negligible impact on the spectral line asymmetry, 
%and have a minor impact of the spectral line asymmetry. }
%hile they affect the determination of the amplitude of the radial velocity curves is reduced by an amount of 3\%.  
Using the spectral-line asymmetry method, we also found the centre-of-mass velocities of \rpup\, and DX~Cet, 
$V_{\mathrm{\gamma}}=47.49\pm0.07$~\kms\, and  $V_{\mathrm{\gamma}}=25.75\pm0.06$~\kms, respectively. 
By comparing our results with previous HARPS observations of classical Cepheids and $\beta$~Cep stars, 
we confirm the linear relation between the atmospheric velocity gradient and the amplitude of the radial velocity curve, 
but {{\it only}} for amplitudes larger than 22.5~\kms. For lower values of the velocity amplitude (i.e., $<22.5$~\kms), our data on $\rho$~Pup seem to indicate that the velocity gradient is null, but this result needs to be confirmed with additional data. We derived the Baade-Wesselink projection factor 
 $p=1.36\pm0.02$ for $\rho$~ Pup and $p=1.39\pm0.02$ for DX~Cet.  We successfully extended the 
 period-projection factor relation from  classical Cepheids to $\delta$~Scuti stars.}
% seem to adhere to the same period-projection factor relation.} }
{}
%{\ep{REALLY NECESSARY?}  These results bring insights into the dynamical structure of pulsating star atmospheres, 
%which help to better understand the k-term problem and the Baade-Wesselink p-factor for Cepheids.}


\titlerunning{$\rho$ Pup and DX~Cet}
\authorrunning{Nardetto et al.}
   \maketitle
%
%________________________________________________________________

\section{Introduction}

The variability of $\rho$ Pup$\equiv$HD\,67523$\equiv$HR\,3185, one of the brightest ($V$=2.88) 
%\citep[$V$=2.88, $b-y$=0.259, $m_1$=0.215, $c_1$=0.731, $\beta$=2.715;][]{kurtz} 
$\delta$ Sct stars, was first reported by \citet{cousins} and then clearly demonstrated by  \citet{eggen} and \citet{ponsen}. 
An accurate value of the  pulsational period ($P=$0.14088143~d) was determined 
from photometric data spanning fifty-two years \citep{moon}.   %It shows a well-defined  pulsation period $P$=0.14088~d. 
Atmospheric parameters (effective temperature, surface gravity, and metallicity) 
were determined from ELODIE spectra by using the MILES library
interpolator \citep{miles}: $T_{\rm eff}$=6810$\pm$121~K, $\log\,g$=3.59$\pm$0.14, and [Fe/H]=+0.60$\pm$0.05.
The [Fe/H] value confirms that \rpup\, is a member of the $\delta$~Del subgroup, that is, late-A and early-F subgiants 
showing spectra with enhanced metal-lines (especially the Fe\,I, Y\,II, Fe\,II and Zr\,II ones) and normal H and Ca\,II lines \citep[][ and references therein]{kurtz}.
\rpup\, is a well-known case where the determinations of the atmospheric parameters by means of Str\"omgren photometry
($b-y$=0.259, $m_1$=0.215, $c_1$=0.731, $\beta$=2.715) disagree in function of
the photometric indices used in the calibrations.  This could be ascribed to its particular metallic content. 
Indeed, the spectroscopic values agree excellently with those obtained from $b-y$ and $c_1$ indices, that is, 6850~K 
and $\log g$=3.5 \citep{breger,kurtz}. Pioneering IUE spectra
show emission features in the Mg~II lines throughout the pulsation cycle
\citep{fraca}.
%Anyway, the values finally adopted by \citet{kurtz} 
%($T_{\rm eff}$=7100$\pm$200~K, $\log\,g$=3.25$\pm$0.20, and [Fe/H]=+0.54) are in good agreement with the spectroscopic ones, 
%also taking into account an intrinsic variability of about 100~K due to the pulsation \citep{yang}. 
Later, detailed 
spectroscopic investigations of the pulsation properties were performed by \citet{mathias97} and \citet{dall}.  
Very recently, the results of a dedicated spectroscopic multisite campaign excluded solar-like oscillations 
with an amplitude per radial mode larger than 0.5~m\,s$^{-1}$ \citep{antoci13}.

DX~Cet$\equiv$HD\,16189 is a bright high-amplitude $\delta$ Sct (HADS) star with a
period $P$=0.103939529~d and full amplitude of 0.20~mag in $V$ light \citep{kiss}.
Stellar parameters were obtained from $uvby\beta$ photometry \citep{kiss,stetson}:
$T_{\rm eff}=7250\pm200$~K, $\log g=3.6\pm0.2$~dex. 

%\citet{dall} considered \rpup\, a radial pulsator and they derived a ratio of 0.43
%%for the ratio between the amplitudes of variations in the H$\alpha$ line and photometry. 
%The spectroscopic parameters are slightly different from those obtained from
%Str\"omgren photometry ($b-y$=0.259, $m_1$=0.215, $c_1$=0.731, $\beta$=2.715), i.e.,
%($b-y$=0.259, $m_1$=0.215, $c_1$=0.731, $\beta$=2.715)$T_{\rm}$=7100~K, $\log\,g$=3.25, and [Fe/H]=+0.54 \citep{kurtz}, but it is probably
%due to the enhancement of the $m_1$ index due to the particular metallic content.
%Indeed, an excellent agreement was obtained just using the 

Comparing the spectroscopic analysis of these two $\delta$ Scuti stars (\rpup\, and DX~Cet) with previous 
results obtained on classical Cepheids and $\beta$~Cep stars is extremely interesting for various reasons. 
First, for a given pulsating star, and also when comparing one star to the other, the spectral lines 
asymmetry show a systematic difference in average (over one pulsation cycle) that is positive or 
negative, depending on the spectral line considered. 
This effect is related to the dynamical structure of the star's atmosphere and has
been investigated in the case of classical Cepheids by means of the k-term  \citep{nardetto08, nardetto09}.
%This effect has been connected to the k-term of 
%Classical Cepheids by \citet{nardetto08, nardetto09}. 
Since the physical origin of this phenomenon is currently not understood,  the comparison
of  the results obtained for several types of pulsating stars 
is particularly helpful. Second, \citet{nardetto13} found a linear relation between the atmospheric 
velocity gradient and the amplitude of the radial velocity curve for eight classical Cepheids and 
two $\beta$~Cep stars ($\alpha$ Lup and $\tau^1$ Lup). This relation is important in the context 
of the Baade-Wesselink projection factor  \citep{nardetto04, nardetto07}, a quantity that is used 
to derive the distance to Milky Way and Large Magellanic Cloud Cepheids \citep{storm11a, storm11b}.  
Improving this relation and generalizing it to other pulsators like $\delta$ Sct 
stars is an interesting roadmap to better understand the Baade-Wesselink projection factor. 
Finally, the use of pulsating stars located below the horizontal branch as distance indicators 
requires { the mode identification as input parameter, and often this is a challenging task to
achieve } \citep{fornax}.  Therefore, 
the full characterization of bright, short-period $\delta$ Sct stars is a key matter in 
calibrating their period-luminosity (P-L) relation.

In Sect. ~\ref{s1} we briefly present our spectroscopic data. 
In Sect.~\ref{s2}, we analyse the pulsating frequencies of the two $\delta$ Scuti stars. 
In Sect.~\ref{s3}, we discuss the implication of our data on the derived distances and on the P-L relations. 
Sect.~\ref{s4} is then devoted to the spectroscopic analysis of the two targets in terms of spectral line asymmetry and velocity.
In Sect.~\ref{s_new}, we compare our results on $\delta$ Sct stars with those on classical Cepheids and $\beta$ Cep stars. 
We close with some conclusions in Sect.~\ref{s5}. 

%%%%%%%%%%
\section{Spectroscopic observations and physical parameters}\label{s1}
%%%%%%%%%%
%High-resolution spectra of \rpup\, and DX~Cet were obtained with HARPS in December 2012 and January 2013.
We observed \rpup\, and \dxcet\, as additional targets in the framework of the ESO LP185.D-0056.
% devoted to CoRoT targets. 
The observations were planned to extend the physical scenario of the $\delta$ Sct stars observed with CoRoT.
%{\bf in the framework of the understanding of the dynamical structure of $\delta$ Sct stars. At this purpose,
%the main requirement was to fully cover one pulsation cycle, not to acquire a dense timeseries.}
  
\begin{figure}[]
   \centering
   \includegraphics[width=\columnwidth,height=\columnwidth]{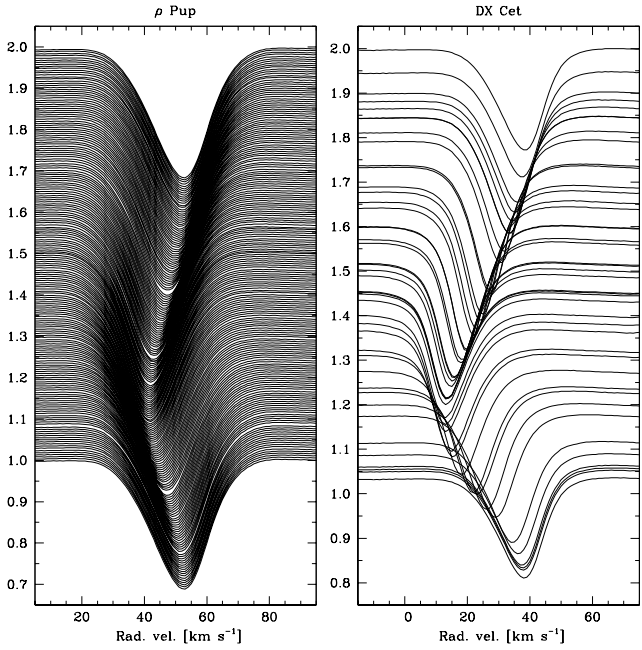}
\caption{ Mean profiles of \rpup\, {\it (left panel)} and
\dxcet\, {\it (right panel)} observed during different nights and folded with the 
respective pulsation period. Spectral intensities  (in continuum units) 
on the y-axis are arbitrarily shifted to clearly show the profile variations.
 }
\label{profmed}
\end{figure}

Three-hundred fifty-nine spectra of \rpup\, were obtained with High-Accuracy Radial-velocity Planetary Searcher 
\citep[HARPS ; ][]{harps}
in the high-resolution mode (HAM, R=115,000)
on five consecutive nights in January 2013.  Observations were performed for a few hours 
at the end of night, as a backup program. The exposure time was set at 30~sec and the
signal-to-noise ratio (S/N) was typically about 270. 
Figure~\ref{profmed} (left panel)  shows the behaviour of the mean line profile of each spectrum folded with the pulsational
period. Forty-three spectra of \dxcet\, were  obtained with HARPS in the high-efficiency mode (EGGS, R=80,000) 
on nine consecutive nights in December 2012.  
Observations were performed for a few hours at the beginning of night, before the rising of  CoRoT targets.
The exposure time was set at 300~sec and the S/N was usually around 200 (see Fig.~\ref{profmed},
right panel). The radial velocity values of \rpup\, were obtained from the HARPS pipeline
\citep{harps}, while those of \dxcet\, were obtained from the mean line profiles
of each spectrum, calculated using the least-squares deconvolution method \citep{lsd}.

%confirms the physical parameters reported above, i.e., 
%$T_{\rm eff}=6650~\pm100$~K, $\log g=3.42\pm0.1$~dex, and [Fe/H]=+0.45$\pm$0.10. 

We estimated the physical parameters of both stars  by means of 
%comparing its spectrum with 
a grid of synthetic spectra. For this purpose
we selected four HARPS spectra with a high S/N that were taken at intervals of  a quarter of a period starting from the  maximum of the radial velocity curve.
%was chosen on the grounds of its high S/N and, 
%more important, distributed over the pulsation cycle.narrow and symmetric lines, i.e., not overly distorted by
%the strong radial pulsation.
%The synthetic spectra were computed from the ATLAS9 models using the ATC
%code \citep[Atmospheric Tools Compilation; ][]{stuetz}. 
%We considered $v_{eq}\,\sin i=7$~\kms. 
To save computing time, we reconstructed the synthetic spectra in five 200~\AA\, segments in the
range 4000-6650~\AA\, by using 
the ATLAS9 models and the ATC code \citep[Atmospheric Tools Compilation; ][]{stuetz}.
% and then we computed the syntetic s
%created several short synthetic spectra within a range of 200~\AA\, 
%in the range 4000-6650~\AA. 
%(4000-4200, 4400-4600, 5000-5200, 5400-5600, and 6000-6200~\AA),
Temperature, gravity, metallicity, and microturbulence velocity were allowed to vary 
in the range  6400 $\leq T_{\rm eff}\leq 8000$~K, $2.6\leq\log~g\leq4.8$~dex (cgs units),
$-1.0\leq$[Fe/H]$\leq1.0$, and $0\leq~v_{mic}\leq6$~\kms, respectively.
%covering Teff=6400 - 8000, logg=2.6 - 4.8, [Fe/H]=-1.0 - 1.0 and vturb=0 - 4 km/s.
 We compared the synthetic 200~\AA-segments with the observed ones by computing the
$\chi^2$. At the end of this process, each segment supplied a set of physical parameters obtained by
averaging its own  best 50 least-squares solutions. In turn, these five sets were averaged
to give the physical parameters at the pulsation phase of the HARPS spectrum that we were considering.
%the  physical parameters were obtained 
% by averaging the best 50 least-squares solutions obtained in the 200~\AA\, range. 
%In such a way,  each of the four HARPS spectra supplied five 
%values of $T_{\rm eff}$, $\log~g$, [Fe/H], and $v_{mic}$. We averaged these five values
%to get both the physical parameters and  their uncertainties at the given four pulsation phases. 
Then, the averages  of the four
$T_{\rm eff}$, $\log~g$, [Fe/H], and $v_{mic}$ values corresponding to the four pulsation
phases supplied the mean stellar parameters.
The typical uncertainties of the parameters supplied by one
spectrum were also adopted as the errorbars of the mean stellar parameters. 
%The standard deviations
%calculated from the four values and the subsequent error on the mean are, instead of the probable error  
%We assumed the typical 

%We then averaged the results for the different ranges.

The procedure described above supplied
$T_{\rm eff}=6650~\pm100$~K, $\log g=3.42\pm0.14$~dex, [Fe/H]=+0.20$\pm$0.06, and 
$v_{mic}=4.2\pm$0.7~\kms\, for \rpup.
%These mean values are in an interval
%of 100~K, 0.02~dex, 0.02~dex, and 0.08~\kms\, in 
%$T_{\rm eff}$, $\log~g$, [Fe/H], and $v_{mic}$, respectively for \rhopup.
Moreover, we measured a projected rotational velocity $v_{eq}\,\sin i=13\pm$1~\kms.
% and a microturbulence velocity $v_t=4.2\pm$0.7~\kms.
In the case of \dxcet\, we obtained 
 %thus determining  
$T_{\rm eff}=7200\pm100$~K, $\log g=3.58\pm0.14$~dex,
$v_{mic}= 3.4\pm0.4$~\kms, [Fe/H]=$-0.27\pm0.06$, and  $v_{eq}\,\sin i=7\pm$1~\kms.
Therefore, our spectroscopic analyses yields out that \dxcet\, has a slight subsolar 
metallic content, while \rpup\, is  confirmed to be  a metal-rich star.

\begin{figure}[]
   \centering
   \includegraphics[width=\columnwidth,height=\columnwidth,angle=270]{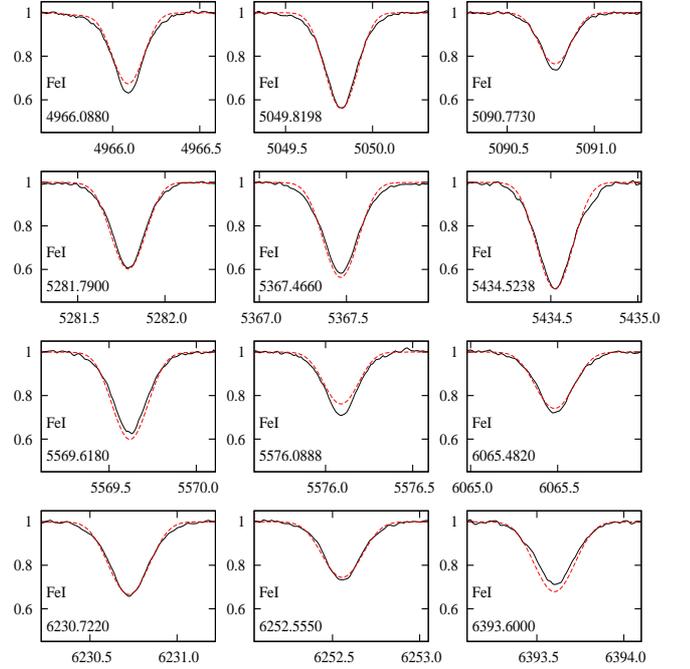}
\caption{Fitting of the FeI lines in the spectrum of \dxcet.
The HARPS spectrum (black solid line) is compared with the synthetic
spectrum obtained with  
$T_{\rm eff}=7100$~K, $\log g=3.55$~dex, $v_{mic}= 3.4$~\kms, and [Fe/H]=$-0.27$ (red dashed).
% and the second with  
%$T_{\rm eff}=7200$~K, $\log g=3.6$~dex, and [Fe/H]=$0.0$ (green dotted and dashed).
Intensities are normalized to the continuum level.}
\label{dxiron}
\end{figure}

To support the  [Fe/H] determination of \dxcet\, in view of
the discussion of the P-L relations (see Sect.~\ref{s3}), Fig.~\ref{dxiron}
shows the agreement between the observed spectrum and the synthetic
one calculated with the above parameters.

%{\bf To clarify better the case of \dxcet\, (see Sect.~\ref{s3}),} Fig.~\ref{dxiron} 
%shows the agreement between the observed spectrum and the synthetic
%one calculated with the above parameters.  
% slightly less than that previously considered. To illustrate this point,
%Fig.~\ref{dxiron} shows the agreement between the observed spectrum and the
%synthetic one with $T_{\rm eff}=....$~K, $\log g=...$~dex, and [Fe/H]=$-0.5$,
%while that with $T_{\rm eff}=7200$~K, $\log g=3.6$~dex, and [Fe/H]=$0.0$
%generally shows lines that are too deep. [TBC!!]}

{\it Hipparcos} parallaxes are available for both stars \citep{hipparcos}: $\pi=8.65\pm0.80$~mas for
DX~Cet and $\pi=51.33\pm0.15$~mas for \rpup. \dxcet\, is located at 116$\pm$5~pc and
we obtain $M_V$=1.68$\pm$0.20 assuming $V$=7.00. \rpup\, is much closer, at 19.48$\pm$0.02~pc
only, and we derive $M_V$=1.36$\pm$0.01 from the well-established $V$=2.81 magnitude.
In the case of \rpup, the  parallax immediately supplied $L=24.0~\pm0.2\,L_{\sun}$.
Combined with the  interferometric determination of the radius
\citep[$R=3.52~\pm0.07\,R_{\sun}$; ][]{antoci13}, this implies 
$T_{\rm eff}=6820\pm170$\,K, in good agreement with our spectroscopic value.
Furthermore, we introduced $T_{\rm eff}$, $P$, $Q$=0.033~d (corresponding to the
fundamental radial mode), $M_V$, and the bolometric correction $+0.03$~mag \citep{torres}
in the  relation \citep{bb} 
\begin{equation}
\log Q = -6.456 + \log P + 0.5 \log g + 0.1 M_{\rm bol} + \log T_{\rm eff},
\label{bb}
\end{equation}
thus obtaining $\log g$=3.70$\pm$0.02. This latter value agrees
within 2$\sigma$ with ours ($\log g$=3.42$\pm$0.14) and   
within 1$\sigma$ with that of \citet{miles} ($\log g$=3.59$\pm$0.14).
When applying the same approach to \dxcet, Eq.~\ref{bb} can only supply the condition
\begin{equation}
\log g + 2.0 \log T_{\rm eff}= 11.56(\pm0.04),
\end{equation}
since we have no radius measurement \citep[see also ][]{exo}. 
Our spectroscopic values match this condition at about 1$\sigma$ (11.30$\pm$0.16).
Since in the case of \rpup\, the $T_{\rm eff}$ value is well constrained, this comparison
shows that the $\log g$ determination is the most critical parameter to evaluate.
% and that an improvement is necessary in its determination when dealing with large amplitude pulsators. 

%%%%%%%%%%
\section{Frequency analyses and nature of pulsations}\label{s2}
%%%%%%%%%%

   \begin{figure}[]
   \centering
   \includegraphics[width=\columnwidth,height=\columnwidth]{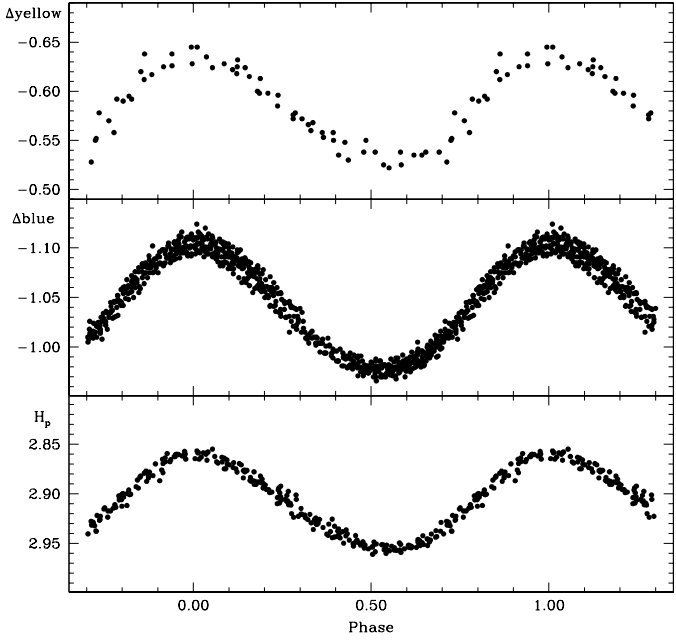}
   \caption{Light curves of $\rho$~Pup. {\it Top panel:} \citet{eggen} data, 60 measurements 
from 19 to 26 March, 1956. {\it Middle panel:}  \citet{ponsen} data, 658 measurements from 
January 30 to April 23, 1961. 
{\it Bottom panel}: {\it Hipparcos} data \citep{esa}, 248 measurements from January 4, 1990 to February 24, 1993 }
              \label{eggen2}
    \end{figure}

 \begin{figure}[]
   \centering
   \includegraphics[width=\columnwidth,height=\columnwidth]{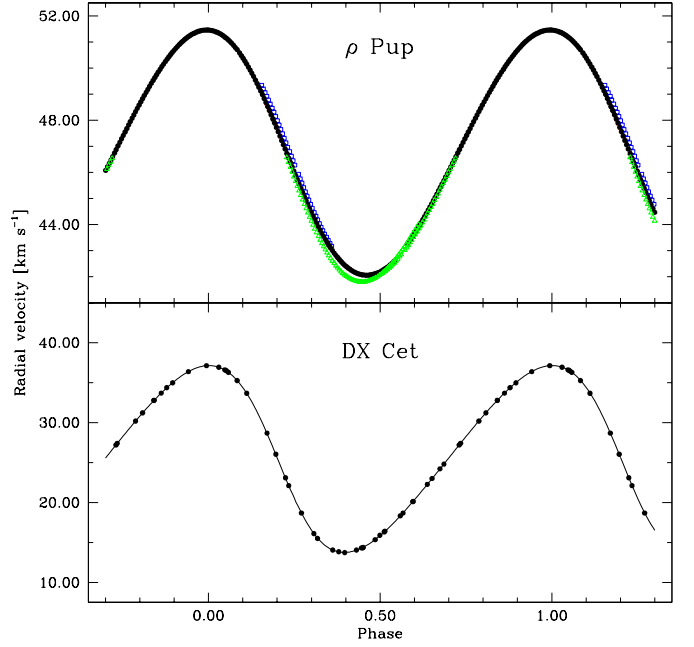}
   \caption{{\it Top panel:} HARPS radial velocity values of $\rho$~Pup. Different symbols (colours) denote different
nights. Cycle-to-cycle variations are clearly visible around the minimum. 
 {\it Bottom panel:}  HARPS radial velocity values of \dxcet. The solid line is the fit with $f_1$ and
three harmonics (Table~\ref{lfit}).}  
              \label{harpsvr}
     \end{figure}
\begin{table*}
\caption{Least-squares parameters of the HARPS radial velocity curves of \rpup\,
(T$_0$=BJD 2456293.0056) and \dxcet\, (T$_0$=BJD 2456272.5889).
}
\begin{tabular}{l rrr c l rrr}
\hline
\hline
\noalign{\smallskip}
\multicolumn{4}{c}{\rpup} && \multicolumn{4}{c}{\dxcet} \\
\cline{1-4} \cline{6-9}
\noalign{\smallskip}
\multicolumn{1}{c}{ID} &
\multicolumn{1}{c}{Frequency} &
\multicolumn{1}{c}{Amplitude} &
\multicolumn{1}{c}{Phase} 
&&
\multicolumn{1}{c}{ID} &
\multicolumn{1}{c}{Frequency} &
\multicolumn{1}{c}{Amplitude} &
\multicolumn{1}{c}{Phase} 
\\
& \multicolumn{1}{c}{[d$^{-1}$]} &
 \multicolumn{1}{c}{[\kms]} &
\multicolumn{1}{c}{[0, 2$\pi$]} 
&&
& \multicolumn{1}{c}{[d$^{-1}$]} &
 \multicolumn{1}{c}{[\kms]} &
\multicolumn{1}{c}{[0, 2$\pi$]} 
\\
\noalign{\smallskip}
\hline
\noalign{\smallskip}
$f_1$    & 7.098 &  4.948  & 0.129                    && $f_1$ & 9.6197 & 11.444 & 0.318 \\
         &$\pm$0.001 & $\pm$0.011 & $\pm$0.008        &&       &$\pm$0.0001 & $\pm$0.007 & $\pm$0.001 \\
$2f_1$   &       &  0.297  & 5.102                    && $2f_1$ &       & 1.890 & 5.308 \\
         &       & $\pm$0.012   & $\pm$0.007          &&        &       &$\pm$0.008   & $\pm$0.005 \\
$3f_1$   &       &  0.039  & 2.541                    && $3f_1$ &       & 0.377 & 3.953 \\
         &       & $\pm$0.012 & $\pm$0.364            &&        &       &$\pm$0.007   & $\pm$0.020 \\ 
$4f_1$   &       &  0.022  & 5.103                    && $4f_1$ &       & 0.092 & 2.542 \\
         &       & $\pm$0.008 & $\pm$0.651            &&        &       &$\pm$0.008   & $\pm$0.089 \\
$f_2$    & 7.900 &  0.281  & 2.613                    && $5f_1$ &       & 0.022 & 0.656 \\
         & $\pm$0.007 & $\pm$0.027 & $\pm$0.149       &&        &       &$\pm$0.008   & $\pm$0.357 \\ 
\noalign{\smallskip}
\multicolumn{2}{l}{Mean rad. vel. [\kms]}&\multicolumn{2}{c}{46.739$\pm$0.002}&&&&\multicolumn{2}{c}{25.516$\pm$0.004}    \\
\multicolumn{2}{l}{Residual r.m.s. [\kms]}&\multicolumn{2}{c}{0.041}&&&& \multicolumn{2}{c}{0.023}  \\
\hline
\label{lfit}
\end{tabular}
\end{table*} 

\subsection{\rpup} \label{f2}%%%%%%%%%%

An exhaustive list of photometric measurements of \rpup\, was reported by 
\cite{moon}. \cite{ponsen}, \cite{eggen}, and {\it Hipparcos} \citep{esa} data are the most
 suitable for a detailed analysis.
% are those obtained by \cite{eggen}, \cite{ponsen}, and the  {\it Hipparcos} ones \citep{tycho}.
%No recent photometric timeseries are available  and those obtained by
%\cite{eggen} and \cite{ponsen} remain the only ones. 
Data were obtained in different passbands:
Ponsen's measurements were taken through a blue filter,
%Eggen's measurements were taken through a yellow filter, 
Eggen's through a yellow filter,
and those of {\it Hipparcos} in the wide $H_p$ passband.
The folded light curves show  asymmetric shapes (Fig.~\ref{eggen2}).
%in clear disagreement with the general statement \citep[e.g., ][]{yang}
The full amplitudes are 0.13, 0.11,  and
0.09~mag, respectively. These values fit  the sequence  observed
at 3858, 4310, 4720, and, 5875~\AA, that is, 0.17, 0.14, 0.12, and 0.09~mag \citep{doss}.
Amplitude ratios ($R_{21}$) and phase shifts ($\phi_{21}$) 
%Fourier parameters 
were obtained from a 
least-squares fit with $f=1/P$  and $2f$ ($P$=0.14088143~d):
$\phi_{21}=3.85\pm0.12$ rad and $R_{21}=0.05\pm0.01$ for Ponsen's blue data,
$\phi_{21}=4.26\pm0.22$ rad and $R_{21}=0.14\pm0.03$ for Eggen's yellow data, and
$\phi_{21}=4.19\pm0.10$ rad and $R_{21}=0.09\pm0.01$ for {\it Hipparcos} data. 
Taking into account the different passbands, these Fourier parameters agree excellently and
suggest significant departure from a perfect sine-shaped light curve.
The Fourier parameters of \rpup\, light curves  are typical
for HADS stars \citep[Figs.~4 and 6 in ][]{ogle}. These stars
pulsate in a radial mode (typically the fundamental one) and nonradial modes have a much smaller
amplitude. However, the amplitude of $\rho$~Pup is not so large that it could be
considered as that of a bona-fide HADS star, and more analyses are necessary.

The multiperiodicity of \rpup\, was evident after 
the detection of clear cycle-to-cycle variations in the radial velocity values obtained with
the Coud\'e Echelle Spectrograph (CES) at the CAT ESO telescope \citep{mathias97}.
%suggested by \citet{yang} and then
%confirmed by \citet{mathias}. The latter authors 
%pointed out clear cycle-to-cycle variations in the radial velocity values obtained with
%the Coud\'e Echelle Spectrograph (CES) at the CAT ESO telescope. 
The main oscillation $f_1$=7.098168~d$^{-1}$ was identified as a radial mode and two nonradial modes with frequency 7.8 and
6.3~d$^{-1}$ were also proposed, but the latter is very uncertain.
We used the iterative sine--wave least--squares fitting method \citep{vani}
to analyse in frequency the radial velocity measurements obtained from 
the HARPS spectra, determined by the on-line pipeline with an internal
error of about 1~m\,s$^{-1}$. After the detection of $f_1$, the power spectrum
clearly shows a structure centred at $2f_1$, confirming that the radial velocity
curve is asymmetrical, as well. The folded radial velocity curve shows both the
asymmetrical shape and cycle-to-cycle variations up to 0.36~\kms\ (Fig.~\ref{harpsvr}). 
Asymmetry can be evaluated from the maximum, which occurs at phase 0.0, while the minimum
occurs before phase 0.50 (black filled circles). 
After introducing $f_1$ and $2f_1$ as known constituents,
the highest peak in the power spectrum of HARPS data is at $f_2$=7.900~d$^{-1}$, in good agreement with the value
obtained from CAT data \citep[i.e., 7.815~d$^{-1}$; ][]{mathias97}. However, we note that the spectral
windows of both datasets
are not adequate to perform a very detailed, self-consistent frequency analysis, since these 
spectroscopic data were
acquired with the goal of studying the dynamics of the dominant pulsation mode.
The ground-based photometric data are too noisy to supply useful hints about $f_2$. 
Even {\it Hipparcos} data are unsuitable to search for a  small amplitude, short-period additional mode.
However, the spectral window of these data is free from the $\pm$1~d$^{-1}$ aliasing, and we could see, among many
others, a peak at 8.860~d$^{-1}$. \citet{antoci13} pointed out 8.82~d$^{-1}$ as a possible additional
frequency. Therefore, we calculated the least-squares fit by considering $f_2$=8.860~d$^{-1}$, but the residual r.m.s.
(0.067~\kms)  is poorer than  that obtained using  $f_2$=7.900~d$^{-1}$ (0.041~\kms). 
An additional analysis of new data  (e.g., Antoci et al. 2013, in preparation) may be able to ascertain the true value.
We adopted $f_2$=7.900~d$^{-1}$ to describe the dynamical structure of \rpup,
 and the effect of this choice is evaluated at the end
of Sect.~\ref{rpup}.

%for the detection of the additional mode(s) excited in \rpup. 
Table~\ref{lfit} lists the least-squares solution obtained from the 
HARPS data.
The mean value of the radial velocity, 46.74~\kms, and the $2K$  value 
(peak-to-peak amplitude of the radial velocity curve), 9.9~\kms, agree
both excellently with all the literature values. The former suggests that \rpup\,
is a single star, the latter that the pulsation is very stable over decades
\citep[see also ][]{moon}. The power spectrum of the residuals shows other modes excited around
$f_1$, but the poor spectral window does not allow us to confirm the values
proposed by \citet{antoci13}. In the other parts of the residual power spectrum 
we were able to observe not only the bunches of peaks related to the higher harmonics of $f_1$,
but also those related to the combination terms between $f_1$ (and harmonics) and
the additional modes. 
%They cannot be unambiguously detected due to the poor spectral window.
As stressed by \citet{antoci13}, these patterns affect the same region in which 
solar-like oscillations are expected to leave their fingerprints. 
%after introducing $f_2$ a known constituent and, amongst them, those related to the harmonics $3f_1$ and $4f_1$. 
%The patterns located around $f_1$ also suggest that other modes are excited, with amplitude of 0.05-0.07~\kms. 
All these unresolved components contribute to increase the residual r.m.s. of the solution.

Neither our radial velocity curve nor Ponsen's photometry (blue light
should be  sensitive to this effect)  show the bump 
provoked by a shock wave crossing the atmosphere \citep{dravins}. 
The  emission features in the  Mg~II lines and the radial velocity curves
in the ultraviolet do not show any particular behaviour either \citep{fraca}.
In the top panel of Fig.~\ref{harpsvr}, cycle-to-cycle variations can be noticed especially  
at the minimum, which seems to occur at different phases. This strange 
behaviour is only apparent because \rpup\, was
observed more at the minimum that at the maximum.
% raises the 
%question if the variations are periodic (i.e., they are due to other modes) or they
%are confined just around the minimum phase, 
For sake of clarity, we investigated the possibility of transient or particular
events in the atmosphere dynamics around maximum compression and subsequent
expansion. 
A close examination of the core of the Ca\,II K line did
not reveal any emission during the observations. Therefore, the cycle-to-cycle variations
in radial velocity data are definitely unrelated to the variations in the chromospheric activity.

%{\bf observed as weak emission features in the Mg~II h- and k-lines all along the pulsation cycle
%\citep{fraca}.}
% On the other hand, IUE spectra
%show emission features in the Mg~II h- and k-lines acquired all along the pulsation cycle
%\citep{fraca}.}  

\citet{dall} investigated the presence of the pulsation in the line indices
of the Balmer series using the DFOSC instrument mounted at the
Danish 1.54m telescope in La Silla.
Indeed, pulsation effects were found, and the authors 
%considered \rpup\, a radial pulsator and they 
calculated a ratio of 0.43 for the ratio between the amplitudes of variations in the H$\alpha$ line and photometry. 
They considered a half-amplitude of 75~mmag from Eggen's measurements, which is that
derived from a simple inspection of the light curve. 
We obtained 0.58 for the yellow data and 0.49 for the blue data from the more accurate values resulting from the least-squares fits. 
Together with the available radial velocity/magnitude amplitude ratio 
\citep[$2K/\Delta m_v$=10.0/0.11=91~\kms~mag$^{-1}$, see ][ for review and summary]{yang}, these results
strengthen the identification of $f_1$ as a radial mode.

\subsection{DX~Cet} %%%%%%%%%%

The solution of the radial velocity curve of \dxcet\, is much simpler than
that of \rpup. The five components $f_1$=9.6197~d$^{-1}$, $2f_1$, $3f_1$, $4f_1$ and $5f_1$
are detected in the power spectrum and provide a very satisfactory
fit of the radial velocity curve (Table~\ref{lfit} and Fig.~\ref{harpsvr}). 
%The same solution can be applied to the 
The fit of the {\it Hipparcos} data \citep{esa} supplied
% The light curve can be fitted with a  $f_1$, $2f_1$, and $3f_1$ series and then 
the Fourier parameters $R_{21}=0.24\pm0.02$, $\phi_{21}=3.9\pm0.1$, and $\phi_{31}=1.4\pm0.4$~rad.
%could be derived. 
They are typical values for HADS stars
\citep[Figs.~4, 5, and 6 in ][]{ogle}. The mean $H_p$ magnitude is 7.077$\pm$0.005,
which in turn yields $V$=7.00 for $B_T-V_T$=0.33 \citep{bessell}. This value and
the full amplitude of 0.21~mag as well agree with those reported by
\citet{kiss}. The $2K/\Delta m_v$=23.1/0.21=110~\kms~mag$^{-1}$ value supports radial
pulsation.
% and the strong asymmetry of light and velocity curves suggests the fundamental
%radial mode, as usual in galactic HADS stars.

%spectral window of the HARPS data did not allow us to obtain a clear indication
%about the excitation of a second mode.
%, in excellent agreement with the ratio expected for radial pulsation (0.50).

%   \begin{figure}
 %  \centering
  % \includegraphics[width=\columnwidth,height=\columnwidth]{Fig4.ps}
  % \caption{$\gamma$-asymmetry and the k-term as a function of the pulsation
%period of $\delta$ Sct stars and Cepheids. OMIT $\beta$ Cep STARS ??  }
  %            \label{kappa}
   % \end{figure}

%%%%%%%%%%
\section{\rpup\, and \dxcet\, as distance indicators}\label{s3}
%%%%%%%%%%
The {\it Hipparcos} parallaxes and the identification of the pulsation modes of \rpup\, and \dxcet\, as the fundamental
radial ones allowed us to test the P-L relations of HADS stars.
There are two P-L relations: 
\begin{equation}
M_V = -1.83 (0.08) - 3.65 (0.07) \log P
\label{free}
\end {equation}
\citep{fornax}, and
\begin{equation}
M_V = -1.27 (0.05) - 2.90 (0.07) \log P -0.19 (0.015) [{\rm Fe/H}] 
\label{feh}
\end {equation}
\citep{mcnamara}.
Eq.~\ref{free}\, was obtained by using all the short-period ($P<0.20$~d)
stars that pulsate in the fundamental radial mode
%HADS and SX Phe (metal deficient $\delta$ Sct stars) known 
and are located in a wide variety of stellar systems. It provides
an excellent fit without knowing any other parameter than the period.   
The metal-dependent term was introduced in Eq.~\ref{feh}\, with the goal to
determine the same  law for HADS and Cepheids and should only be considered when
[Fe/H]$<0.0$ \citep{mcnamara}.

For \dxcet, Eq.~\ref{free} supplies $M_V=1.76\pm0.10$ 
%, in good agreement with the {\it Hipparcos} parallax. 
and Eq.~\ref{feh}\, $M_V=1.64\pm0.07$ 
%$M_V=1.60\pm0.07$ for [Fe/H]=0.0 and $M_V=1.69\pm0.07$ 
for [Fe/H]=$-0.27$. Both predictions agree excellently with the 
{\it Hipparcos} $M_V$, and hence $P$=0.104~d can definitely be 
identified as the fundamental radial mode. 
%A higher overtone would not reconcile the {\it Hipparcos} $M_V$ and  the P-L relations.
For instance, by assuming that the pulsation is in the first overtone mode, the corresponding fundamental
mode would be 0.135~d, and the P-L relations would predict $M_V$=1.3, which does 
match the {\it Hipparcos} $M_V$. 
Due to the relevant error on the parallax, \citet{mcnamara} 
applied an LK correction \citep[$-0.10$~mag, then $M_V$=1.58$\pm$0.20; ][]{lk}, but
the star becomes too bright to fit the value predicted by Eq.~\ref{feh} in a more satisfactory way.  
%The improvement in the agreement with 

%but with the new [Fe/H] value the agreement
%between the predicted values  
%and the modified {\it Hipparcos} value ($M_V$=1.58$\pm$0.20).
%becomes less good.}
%\citep[--0.10~mag; ][]{lk}, as suggested by \citet{mcnamara} due to 
%relevant error on the parallax.} 
%Regarding \dxcet, the two relations supply $M_V=1.76\pm0.10$ and $M_V=1.60\pm0.07$, 
%respectively, and both fit the {\it Hipparcos} $M_V$ value. The star has a solar metallic
%content and then the [Fe/H] correction has no impact. The agreement between
%Eq.~\ref{feh}\, and the {\it Hipparcos} $M_V=1.68\pm0.20$ values improved when introducing the
%LK correction \citep{lk}.
%Such a  correction has to be applied when the error on the parallax is relevant 
%and amounts to $-0.10$~mag for \dxcet.  Hence,
%we get $M_V$=1.58$\pm$0.20. 
However, because it is a statistical procedure, 
the LK correction is powerful when applied to a sample of stars (with same $\pi$) 
and is much more uncertain when applied to a single case \citep[e.g., ][]{oud}.

%,,Eq.~\ref{feh}\, is in better agreement with this corrected value,
%but Eq.~\ref{free}\, is still fitting it within respective error bars. 

The period of \rpup\, is among the longest in the $\delta$ Sct sample. Eqs~\ref{free}\,
and  Eq.~\ref{feh}\, predict similar values without the metal-dependent term, that is,  $M_V=1.28\pm0.10$
and $M_V=1.21\pm0.07$, respectively. The former agrees excellently with 
the very accurate {\it Hipparcos} $M_V=1.36\pm0.01$ value, while the latter is 
slightly outside errorbars. We note that using of the [Fe/H] correction in  Eq.~\ref{feh}
shifts the calculated value too far ($M_V=1.09\pm0.07$) from the observed one.
Therefore, this correction definitely does not have to be applied to stars
when [Fe/H]$>0.0$. Moreover, assuming that $f_1$ is the first radial overtone, we 
derive $M_V$=0.87 from  the P-L relations, a too bright luminosity for the
{\it Hipparcos} parallax. This convinced us that \rpup\, is pulsating in
the fundamental radial mode.

This achievement adds an important stellar candle to the sample of galactic
$\delta$~Sct stars with known parallaxes. The current inventory is composed of
nine stars \citep[Table~2 in ][]{mcnamara}. However, the identification of the 
fundamental radial mode is very reliable only for high-amplitude
pulsation, that is, the HADS stars DX~Cet, AI Vel, and SX Phe.
%\citep[Table~2 in ][]{mcnamara} includes only three well-established 
%fundamental radial mode pulsators, i.e., the HADS stars  
%DX Cet, AI Vel and SX Phe. 
One can add $\delta$~Sct itself, since detailed
investigations \citep{templeton} corrected the previous identification of
the dominant mode as a radial overtone \citep{balonadelta}. 
The remaining five cases are rather uncertain, starting with
1~Mon, whose dominant mode  has been identified as an overtone \citep{balonamono}.
It is instructive to note that the dominant modes of FG Vir and
X Cae are nonradial \citep{zima,xcae}:
the fundamental radial mode, if it is indeed excited, has to be
searched for in the low-amplitude terms, with a wide margin of
uncertainty. 
%that their typifing as radial modes is not immediate and misleading.   
%In such a context, the assumption that the dominant mode of 1~Mon is the fundamental one is
%not justified, since it has been identified as an overtone \citep{balonamono}.
%On the other hand,
%the identification of the main pulsation mode of 1~Mon as a radial overtone 
%\citep{balonamono} is still the only available.  
%The dominant mode $f$=12.72~d$^{-1}$ of FG Vir is a nonradial one and  
%the identification of $f$=12.15~d$^{-1}$ as the fundamental mode  
%is still unclear \citep{zima}. Same situation  for X~Cae, where the dominant mode $f=$7.37~d$^{-1}$
%is a nonradial one and the identification of the fundamental one is unclear
%\citep{xcae}. 
$\alpha$~Aql (Altair) is a fast rotating star 
and the nature of the very weak amplitude variations ($<$0.001~mag) 
detected with the WIRE satellite can be generated by a wide variety of modes.
In particular, there is no convincing evidence that the observed modes are radial ones 
\citep{altair}.  Finally, no
mode identification has been  proposed for the small-amplitude variations
(0.02~mag) observed for BS~Tuc. Therefore, \rpup\, is by far the most convincing case of a 
% low-amplitude 
$\delta$~Sct pulsating in the fundamental radial mode
%reliable mode identification among low-amplitude pulsators 
with an accurate {\it Hipparcos} parallax.

%%%%%%%%%%
\section{Spectroscopic analysis}\label{s4}
%%%%%%%%%%

 \begin{figure}
   \centering
   \includegraphics[width=\columnwidth,height=\columnwidth]{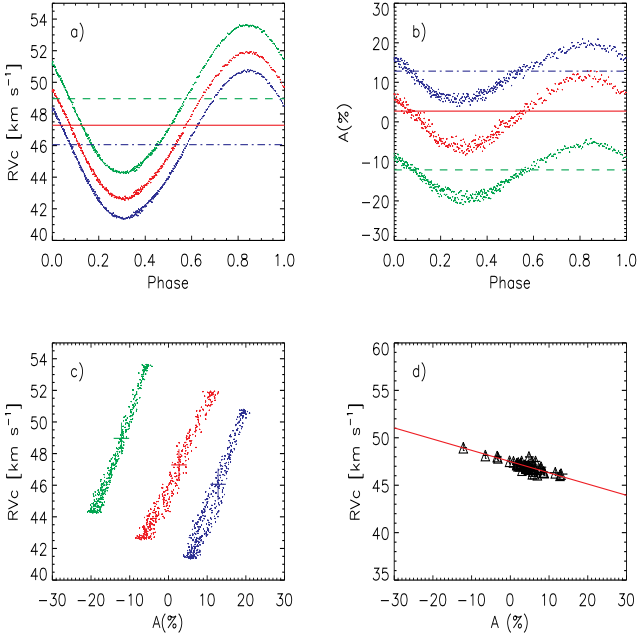}
   \caption{Illustration of the method applied to derive the centre-of-mass velocity of \rpup\, from the line asymmetries. 
Points are single observations folded with $f_1$=7.098~d$^{-1}$; the contribution from $f_2$=7.900~d$^{-1}$
has been removed. }
              \label{ref_fig_rpup}
    \end{figure}

   \begin{figure}
   \centering
   \includegraphics[width=\columnwidth,height=\columnwidth]{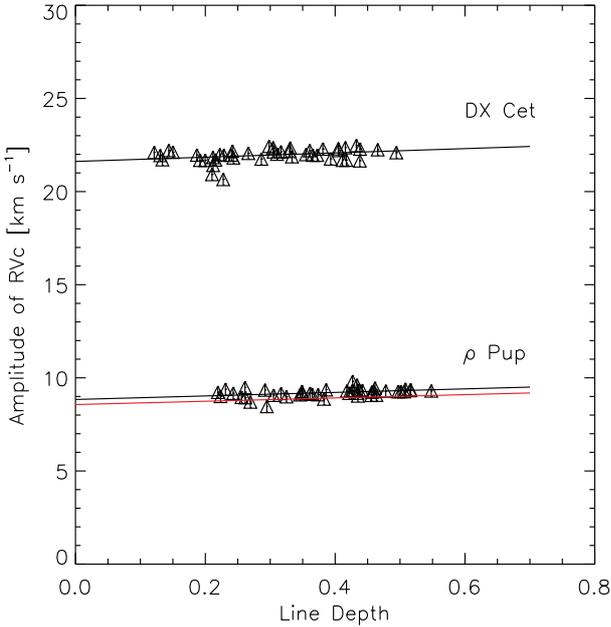}
   \caption{Amplitude of the $RV_{\mathrm{c}}$ curves as a function of the spectral line depth in the case of the two $\delta$ Sct stars \rpup\ (lower values), and \dxcet~(larger values). In the case of $\rho$ Pup, red solid line show the same relation but after removing the nonradial mode. }
              \label{depth}
    \end{figure}

We performed a detailed analysis of the high-S/N HARPS spectra of \rpup\, and
\dxcet\, following the procedure described in \citet{nardetto06}. We selected 45 spectral lines,
taking care that they were not blended. 

We extracted radial velocity and line asymmetry curves for all selected lines of all stars. The method we use for the radial velocity (that is the best one to use when the signal-to-noise ratio allows it) is the first moment of the spectral line, RV$_\mathrm{c}$. The radial velocity curve derived from this method is absolutely independent of the spectral line width and the rotation. This property is extremely valuable for comparing the behaviour of different spectral lines of different pulsating stars.  We also derived the spectral line asymmetries $A$ with a very high precision, using an estimator that we call the bi-Gaussian: two analytic semi-Gaussians are fitted to the blue and red part of the spectral line profile. The amount of asymmetry (in percentage) is then given by the comparison of the half-width at half-maximum of each semi-Gaussian (see Nardetto et al. 2006a, their Eqs. 2 and 3). This definition was well-suited to the data quality. 

%{\it TABLE WITH ALL VALUES, as Tab. 2 for $\alpha$ Lup?}

%{\bf Observed values RV$_\mathrm{c}$ and $A$ are fitted by means of the Fourier series }

\subsection{\rpup} \label{rpup}%%%%%%%%%%

We performed a least-squares fit of the observed values of $RV_\mathrm{c}$  and $A$ by means of the Fourier series

\begin{equation}
y(t)=B_0 + \sum_{i=1}^5 B_i \cos((t-T_0)\,\,f_i + \phi_i) 
\label{fourier}
\end{equation}

and by using $f_1$=7.098d$^{-1}$, $f_2$=7.900~d$^{-1}$, $f_3=2f_1$,  $f_4=3f_1$, and $f_5=4f_1$. In this way we were able to fold the data on one mode by subtracting the contribution of the other. 
Figure~\ref{ref_fig_rpup} shows the folded points with $f_1$ of the radial velocities (panel {\it a}) and asymmetries (panel {\it b}) for the three following lines, \ion{CrII} 4634.07 (blue), \ion{FeI} 5410.91 (green), and \ion{FeI} 6393.60~\AA\ (red), respectively. The corresponding averages of these curves (or similarly $B_0$ values obtained from Eq.~\ref{fourier}) are shown as horizontal lines in panels {\it a} and {\it b} (in dash-dotted, dashed and solid lines), and are called $\gamma$-velocities and $\gamma$-asymmetries. The residual systematic scatter is caused by the contamination from very small amplitude modes that are still not identified after $f_1$ and $f_2$. There is an evident anticorrelation: higher velocities correspond to negative asymmetries, low velocities to positive asymmetries. This behaviour is also shown by the trend in the barycenters (big crosses) of the $RV_\mathrm{c}-A$ loops (Fig.~\ref{ref_fig_rpup}, panel~{\it c}).  By plotting for each line the $\gamma$-velocity  and the $\gamma$-asymmetry, we can calculate the parameters of the analytic relation  
\begin{equation}
V_\gamma(i)=\alpha_0\,A_\gamma(i) + \beta_0, 
\label{quattro}
\end{equation}
thus obtaining $\alpha_0=-0.12\pm0.01$~\kms\, and $\beta_0=47.49\pm0.07$~\kms\,
(Fig.~\ref{ref_fig_rpup}, panel {\it d}). Following \citet{nardetto08}, the centre-of-mass 
velocity of the star is defined as $V_{\gamma\star} =\beta_0=47.49\pm0.07$~\kms 
(i.e., the $\gamma$-velocity corresponding to a null $\gamma$-asymmetry). 
We note that the contributions of $f_2$=7.900~d$^{-1}$ to the $\gamma$-velocity 
and $\gamma$-asymmetry are much less important than those of $f_1$, which means that they affect the spectral line asymmetries
in a marginal way.
%negligible, which seems to indicate that the nonradial mode of $\rho$ Pup is not affecting the 
%spectral line asymmetries (at least in average, i.e. over one pulsating cycle). 
The k-term, defined as k$= <V_{\gamma}>_{\mathrm{i}} - V_{\gamma\star} = 47.0_{\pm 0.1} - 47.5_{\pm 0.1} = - 0.5{\pm 0.1}~$\kms, where $<V_{\gamma}>_{\mathrm{i}}$ is the $\gamma$-velocity averaged over the 45 spectral lines. We briefly discuss the k-term problem in Sect. 5.3.

The next step of the spectroscopic analysis is to measure the velocity gradient within the atmosphere of the star. In \citet{nardetto07}, we have shown that the line depth (taken at the minimum radius phase) is a good indicator of the line-forming regions. We use this definition of the line depth (hereafter $D$) in the following. In this case, the photosphere corresponds to a null line depth. By comparing the 2K amplitude (defined as the amplitude of the first-moment radial velocity curve, hereafter $ \Delta RV_{\mathrm{c}}$) with the depth of the 45 spectral lines selected, one can in principle directly measure the atmospheric velocity gradient (at least in the part of the atmosphere where the lines form). For $\rho$~Pup, we found the following relation (see also Fig.~\ref{depth}): 

\begin{equation}
\Delta RV_{\mathrm{c}} = [0.95 \pm 0.49] D + [8.84 \pm 0.19] \:\mathrm{km\: s}^{-1} \, .
\label{cinque}
\end{equation}

This relation (in which we define the slope and the zero-point by $a_0$ and $b_0$, respectively) can be used 
to derive the Baade-Wesselink projection factor following the semi-theoretical approach presented 
in \citet{guiglion13}. The projection factor is used to convert the radial velocity into the pulsation 
velocity in the Baade-Wesselink methods of distance determination. It is composed of three terms: 
$ p = p_\mathrm{0} f_\mathrm{grad} f_\mathrm{o-g}$ \citep{nardetto07}.  First, the geometric 
projection factor, $p_\mathrm{0}$, is mainly related to the limb-darkening of the star.  
The linear limb-darkening law of the continuum intensity profile of the star provided 
by \citet{claret11} is $I(\cos(\theta))=1-u_{\mathrm R}+u_{\mathrm R}\cos(\theta)$, 
where $u_{\mathrm R}$ is the limb darkening of the star in $R$ band and $\theta$ is the 
angle between the normal of the star and the line of sight. 
For \rpup\, we found $u_{\mathrm R}=0.484$ (considering the rounded values $T_{\mathrm{eff}}=6500$K, $\log g=3.5$, 
$v_{mic}$=4~\kms, and $Z=0.0$). 
Using the relation linking $p_\mathrm{0}$ to 
$u_{\mathrm R}$: $p_\mathrm{0}=\frac{3}{2}-\frac{u_{\mathrm R}}{6}$ \citep{getting34, hadrava09}, 
we found $p_\mathrm{0}=1.416$. In this determination of the geometric projection factor, 
we assumed (as done previously for Cepheids) that the limb-darkening variation within the lines \citep{nardetto07} and that the geometric projection factor time-dependency in the R band \citep{nardetto06b} are negligible. 

Then, $f_\mathrm{grad} $ depends on the  spectral line considered: $f_{\mathrm{grad}} = b_0 / (a_0D+b_0)$ (Nardetto et al. 2007, their Eq.~3). For $\rho$ Pup, it ranges from 0.94 ($D=0.22$) to 0.98 ($D=0.55$). For a typical depth in our sample of $D=0.4$, we find $f_{\mathrm{grad}}=0.96\pm0.01$. 
%This value can be considered as consistent with the use of the cross-correlation method of the radial velocity determination,  which is indeed using a template including many spectral lines.  
Finally, there is a linear relation between $f_{\mathrm{o-g}}$ (correction to the projection 
factor due to the differential velocity between the \emph{optical} and \emph{gas} layers at the  
photosphere of the star) and $\log P$: $f_{\mathrm{o-g}} = [-0.023 \pm 0.005]\,\log{P} + [0.979 \pm 0.005]$ 
\citep[derived using classical Cepheids;][]{nardetto07}.  
When we apply this relation for \rpup\, as in \citet{guiglion13} for the 
$\delta$~Sct stars AI Vel and $\beta$ Cas, we found $f_{\mathrm{o-g}}=1.00 \pm 0.02$. 
The final value of the projection factor of \rpup\, is thus $ p = p_\mathrm{0} f_\mathrm{grad} f_\mathrm{o-g} = 1.36\pm0.02 $. 

From a methodological point of view, it is interesting to note that 
we find $a_0=0.88\pm0.49$ and $b_0=8.56\pm0.19$
(i.e., a decrease of the slope and zero-point of the relation by 8\% and 3\%, respectively) 
when fitting the original data with $f_1$ and harmonics alone.
%\ep{considering} the nonradial mode $f_2$. 
In turn, this increases the value of $f_\mathrm{grad}$, and
consequently that of $p$, by only  0.2\%.
In a similar way, there is no change in the parameters of Eq.~\ref{cinque}
and a non-significant  decrease of the zeropoint of Eq.~\ref{quattro} 
(from 8.86 to 8.84~\kms) when using $f_2=8.86$~d$^{-1}$ instead of 7.90~d$^{-1}$
(see Sect.~\ref{f2}).

%Interestingly, when removing the nonradial mode ($f_2$=7.900~d$^{-1}$), we find $a_0=0.88\pm0.49$ and $b_0=8.56\pm0.19$ (i.e. a decrease of the slope and zero-point of the relation by 8\% and 3\%, respectively). This increases however the value of $f_\mathrm{grad} $ by only 0.2\% (which corresponds to a decrease of the projection factor of 0.2\%).

\subsection{DX~Cet} %%%%%%%%%

\begin{figure}
   \centering
   \includegraphics[width=\columnwidth,height=\columnwidth]{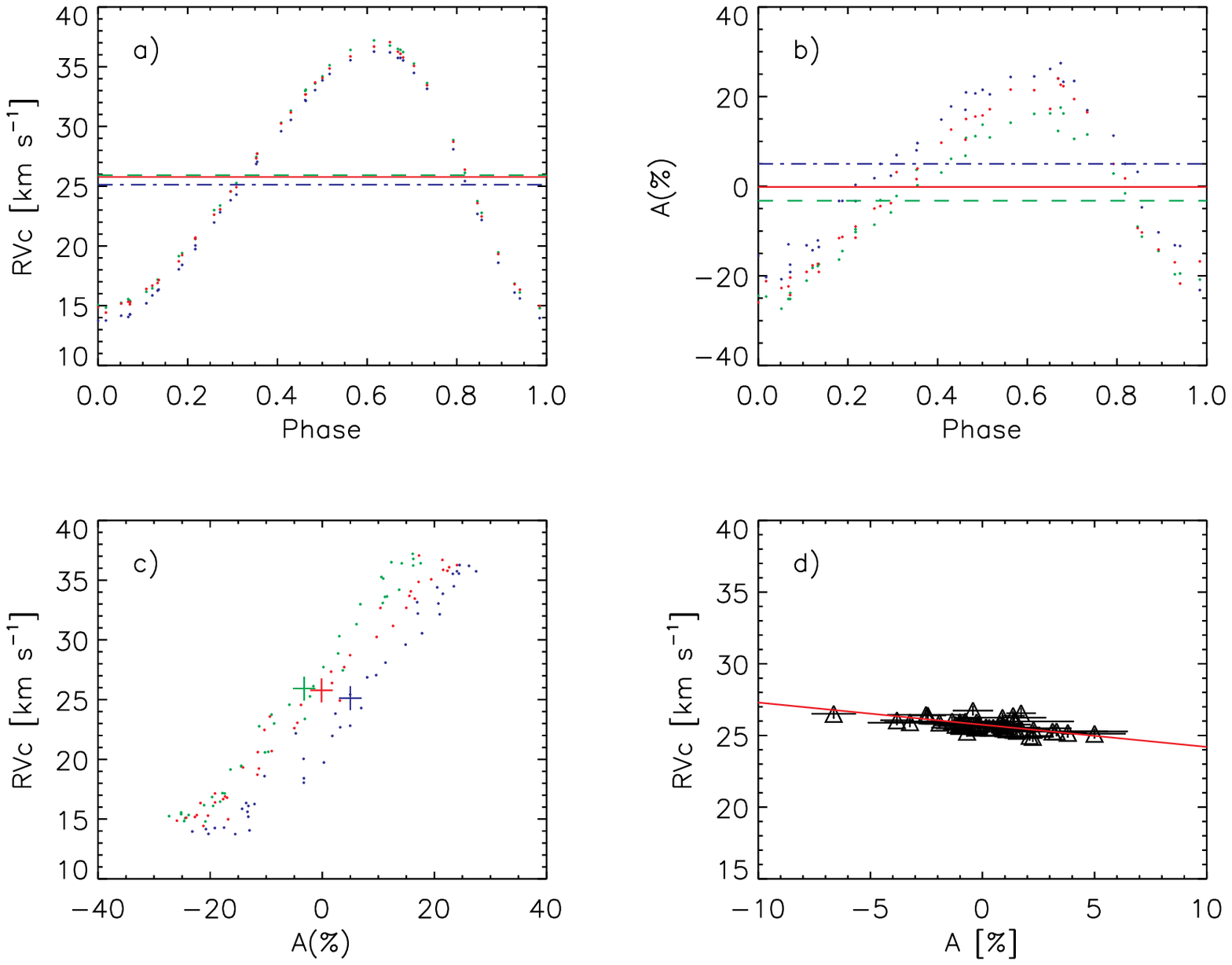}
   \caption{Same as Fig.~\ref{ref_fig_rpup} for \dxcet. }
              \label{ref_fig_dxcet}
    \end{figure}

We applied the same analysis to DX~Cet. Figure~\ref{ref_fig_dxcet} (panels {\it a} and {\it b}) 
shows the radial velocity and line asymmetry curves for the three lines 
\ion{FeI} 4213.65, \ion{FeI} 5393.17, and \ion{FeI} 6393.60~\AA.  We found the same anticorrelation between the $\gamma$-velocity and the $\gamma$-asymmetry: 
$$V_\gamma(i)= [ -0.15\pm 0.03]  A_\gamma(i) +  [25.75\pm 0.06].$$ 

The centre-of-mass velocity of the star is then $V_{\gamma\star} =\beta_0=25.75\pm0.06$  \kms and the k-term is 
$k =  <V_{\gamma}> - V_{\gamma\star} = 25.7_{\pm 0.1} - 25.7_{\pm 0.1}  \simeq 0.0{\pm 0.1}  \:\mathrm{km\: s}^{-1} $.

For the atmospheric velocity gradient we found (Fig. \ref{depth})
\begin{equation}
\Delta RV_{\mathrm{c}} = [1.15 \pm 0.48] D + [21.62 \pm 0.14] \:\mathrm{km\: s}^{-1},
\end{equation}
which corresponds to a velocity gradient or $f_\mathrm{grad}$ correction of  the projection factor ranging 
from 0.97 ($D=0.12$) to 0.99 ($D=0.49$). The average value $f_\mathrm{grad}=0.98\pm0.01$ ($D=0.30$) was used. 
%With $T_{\mathrm{eff}}=7250$K,  $\log g=3.5$, $v_t=2$ \kms and the solar metallicity, we obtain $u_{\mathrm V}=0.49$ and a geometric projection factor of $p_\mathrm{0}=1.42$. Considering, as for $\rho$ Pup, $f_{\mathrm{o-g}}=1.00 \pm 0.02$, we finally find a Baade-Wesselink projection factor for DX Cet of $p=1.39\pm0.02$.
From the rounded values available in  \citet[][  
$T_{\mathrm{eff}}=7000~$K,  $\log g=3.5$, $v_{mic}=2.0$~\kms\, and Z=$-0.3$]{claret11},
%Considering the following available rounded values in the Table of  \citet{claret11}, $T_{\mathrm{eff}}=7000~$K,  $\log g=3.5$, $v_{mic}=2.0$~\kms\, and Z=--0.30, 
we obtained $u_{\mathrm R}=0.474$ and a geometric projection factor $p_\mathrm{0}=1.421$. 
Considering, as for \rpup,  $f_{\mathrm{o-g}}=1.00 \pm 0.02$, we finally 
found a Baade-Wesselink projection factor $p=1.39\pm0.02$ for \dxcet.

\section{Comparing $\delta$~Scuti stars with $\beta$ Cep stars and classical Cepheids}\label{s_new} %%%%%%%%%%

%We included in our HARPS sample stars with
Our HARPS sample is composed of stars with 
a good phase coverage and for which we were able to measure the spectral line asymmetry properly.
We  have eight classical Cepheids \citep[R~TrA, S~Cru, Y~Sgr, $\zeta$ Gem, $\beta$~Dor, RZ~Vel, $\ell$~Car, and RS~Pup; ][]
{nardetto08}, two $\delta$~Sct stars  (\rpup\, and \dxcet; this paper), and the $\beta$~Cep star $\alpha$~Lup 
\citep{nardetto13}. Moreover, we have other cases.
%We remind that the 
AI~Vel has an incomplete phase coverage that prevented us from measuring the $\gamma$-asymmetry, although  we were able to measure 
the atmospheric velocity gradient and the projection factor.  $\beta$~Cas  is
rotating too fast to measure the spectral line asymmetry or derive the atmospheric velocity gradient 
with a good precision. Thus,  we determined the projection factor only  with  
a large uncertainty \citep{guiglion13}.
The classical Cepheid X~Sgr shows spectral line splitting due probably 
to a shockwave in the atmosphere \citep{mathias06}. The $\beta$~Cep $\tau^1$~Lup shows significant
nonradial modes \citep{nardetto13} and we were unable to consistently apply the bi-Gaussian method  
to measure the line asymmetry.

%Our sample of pulsating stars observed with HARPS is now quite significant. If we include only stars with a good phase coverage and for which we could measure the spectral line asymmetry properly, we have eight classical Cepheids: R~TrA, S~Cru, Y~Sgr, $\zeta$ Gem, $\beta$~Dor, RZ~Vel, $\ell$~Car, and RS~Pup \citep{nardetto08}, two $\delta$~Sct stars (\rpup\, and \dxcet, this paper), and one $\beta$~Cep star $\alpha$~Lup \citep{nardetto13}, which corresponds to eleven stars in total. We remind that the $\delta$~Sct star AI~Vel presented in \citet{guiglion13} had an incomplete phase coverage which prevented us to measure the $\gamma$-asymmetry (but we could measure the atmospheric velocity gradient and the projection factor), while $\beta$~Cas was rotating too fast to measure the spectral line asymmetry or derive the atmospheric velocity gradient with a good precision (we could however estimate the projection factor, but with a large uncertainty). Similarly, in the case of the classical Cepheid X~Sgr \citep[which shows spectral line splitting due probably to a shockwave in the atmosphere,][]{mathias06} and for the $\beta$~Cep $\tau^1$~Lup \citep[which has  non-negligible nonradial modes, ][]{nardetto13}, we could not consistently apply the bi-Gaussian method  to measure the line asymmetry. }

   \begin{figure}
   \centering
   \includegraphics[width=\columnwidth,height=\columnwidth]{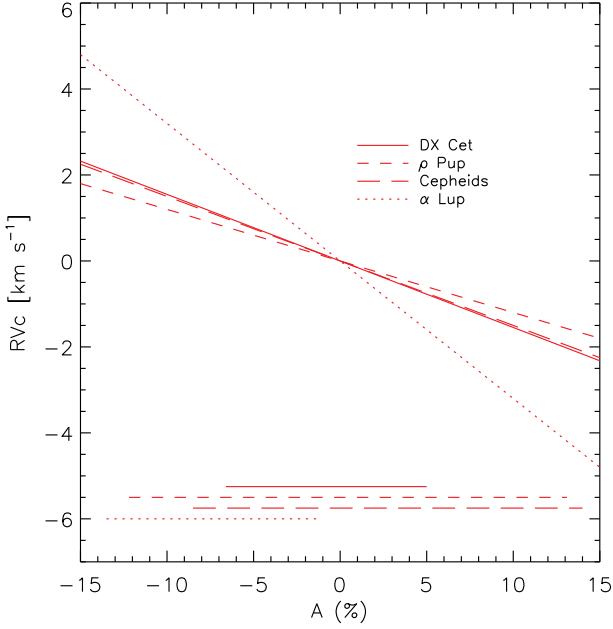}
   \caption{Same as Figs.~\ref{ref_fig_rpup}d and  \ref{ref_fig_dxcet}d, but including the results obtained for classical Cepheids (averaged over eight stars) and $\alpha$ Lup ($\beta$~Cep star). The horizontal lines at the bottom 
show the corresponding $\gamma$-asymmetry ranges.}
              \label{a0}
    \end{figure}
    
   \begin{figure}
   \centering
   \includegraphics[width=\columnwidth,height=\columnwidth]{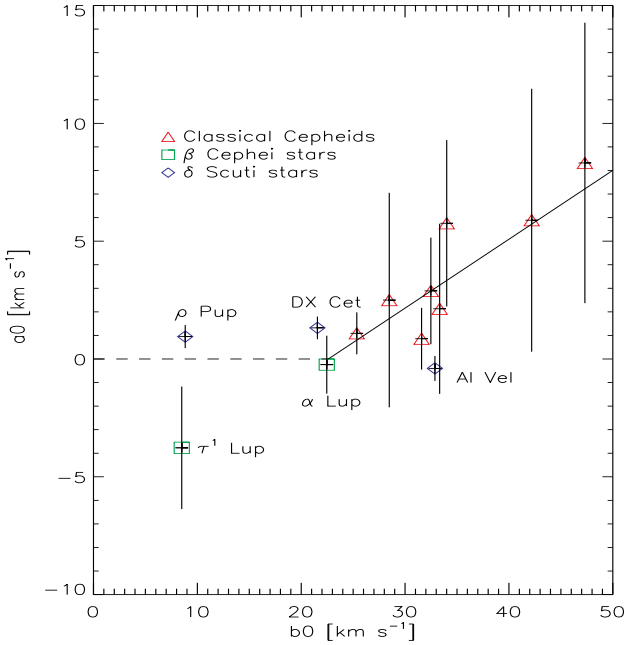}
   \caption{Velocity gradient in the atmosphere of the stars as a function of the amplitude of the $RV_{\mathrm{c}}$ curves.}
              \label{velgrad}
    \end{figure}

  \begin{figure}
   \centering
   \includegraphics[width=\columnwidth,height=\columnwidth]{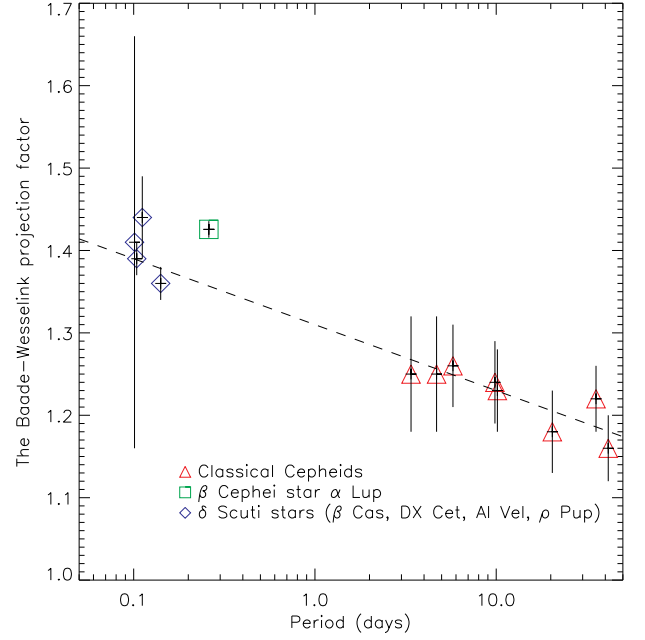}
   \caption{The Baade-Wesselink projection factor as a function of the period for different kinds of pulsating stars. The $\delta$~Scuti stars indicated as blue diamonds are, by increasing period: $\beta$~Cas, DX~Cet, AI~Vel and $\rho$~Pup.}
              \label{Pp_pulsating}
    \end{figure}

In Fig.~\ref{a0}, we compare $V_\gamma(i)=\alpha_0\,A_\gamma(i) + \beta_0$ relations between 
the $\gamma$-velocities and the $\gamma$-asymmetries  (after correcting for the zero-point)  found for $\rho$~Pup and DX~Cet, with previous 
results obtained for the $\beta$ Cep star $\alpha$~Lup and for the eight classical Cepheids mentioned above. 
It is remarkable that the results obtained for $\delta$~Scuti stars and classical Cepheids are consistent, while $\alpha$~Lup shows a different behaviour (steeper slope of the relation). We also note that $\alpha$~Lup has only negative $\gamma$-asymmetries \citep[at least over the 55 spectral lines considered in ][]{nardetto13}. It is currently very 
difficult to interpret this behaviour, but at least qualitatively, we can conclude that there is a  
particular physical mechanism that probably affects the $\alpha$~Lup spectral lines asymmetry, 
while this is not the case for the $\delta$~Scuti and Cepheid stars. 

For the k-term quantity, we found $k=-0.5$~\kms\, and $k=0$~\kms\, for $\rho$~Pup and DX~Cet, respectively. 
These values are consistent with those found for Cepheids 
\citep[values ranging from 0 to $-1$~\kms\, depending on 
the period of the star, ][]{nardetto13}. By using a Cepheid in an eclipsing binary system ($P_{\mathrm{puls}}=3.8$ d, $P_{\mathrm{orb}}=309$ d), \citet{pilecki13} independently found a blue-shifted value k=$-0.59~$\kms. 
On the other hand, $\alpha$~Lup shows a red-shifted value k=2.2~\kms. 

% {\bf It is worth noticing that the $V_\gamma(i)=\alpha_0\,A_\gamma(i) + \beta_0$ relation and the k-term values are probably physically connected.}

Figure~\ref{velgrad} shows the slope $a_\mathrm{0}$ (from the $\Delta RV_{\mathrm{c}} = a_\mathrm{0} D + b_\mathrm{0}$ relation) as a function of the 2K-amplitude (i.e.,  $b_\mathrm{0}$). The results obtained for $\delta$~Scuti stars, $\beta$ Cep stars, and classical Cepheids are compared. The case of $\tau^1$~Lup must be considered separately, as previously mentioned in  \citet{nardetto13}. This star indeed shows a reverse atmospheric velocity gradient, which means that a line forming in the upper part of the atmosphere has a 2K-velocity amplitude lower than a line forming closer to the photosphere. This might be because of the nonradial mode detected in the spectroscopic data. On the other hand, $\rho$~Pup has a null velocity gradient for a similar velocity amplitude of about $b_\mathrm{0} \simeq 10$ \kms. It has to be confirmed, but it seems that stars with a dominant radial mode and with $b_\mathrm{0}$ lower than $22.5$~\kms\, have a null atmospheric velocity gradient (dashed line in the figure). Conversely, we can convincingly consider a linear trend (solid line in the figure) for 2K-amplitudes larger than 22.5~\kms: $a_\mathrm{0} = [0.29\pm0.04] b_\mathrm{0} - [6.61\pm1.36]$.

The comparison of the projection factors within our sample is also  very interesting.
%We compare the projection factor for the pulsating stars in our sample. 
In the framework of our project, we determined the Baade-Wesselink 
projection factor for four $\delta$~Sct stars: \rpup\, ($p=1.36\pm0.02$), 
\dxcet\, ($p=1.39\pm0.02$), AI~Vel  ($p=1.44\pm0.05$), and $\beta$~Cas ($p=1.41\pm0.25$).  
Figure~\ref{Pp_pulsating} shows that all 
these values excellently fit the extension toward short periods of the relation found for Cepheids, that is, $p = [-0.08 \pm 0.05] \log P + [1.31 \pm 0.06]$ \citep{nardetto09}. 
This result seems more robust than the similar one obtained by \citet{laney09}, who used an indirect method
based on the comparison of geometric and pulsation parallaxes.
On the other hand,
%These values seem to be consistent with the period-projection factor relation found for Cepheids (dashed line in Fig. \ref{Pp_pulsating}): $p = [-0.08 \pm 0.05] \log P + [1.31 \pm 0.06]$ \citep{nardetto09}.
the  projection factor  of the $\beta$~Cep star $\alpha$~Lup is 8$\sigma$ higher than the relation 
(Fig.~\ref{Pp_pulsating}). By omitting $\alpha$ Lup, we can determine the following relation common to
$\delta$ Sct stars and classical Cepheids: 

\begin{equation}
p = [-0.08 \pm 0.01] \log P + [1.31 \pm 0.01], 
\end{equation}
which is similar to the one derived for Cepheids only, but more precise (the reduced $\chi^2$ is 0.97).  

%the $\beta$~Cep star $\alpha$~Lup has a projection factor at 8$\sigma$ above the relation (Fig.~\ref{Pp_pulsating}).}
%find also that our values are consistent with the ones found by \citet{laney09} for another sample of $\delta$~scuti stars using an indirect method based on the \emph{PL} relation of Cepheids and $\delta$~Scuti stars. }
 
However, we have to stress some methodological points. First, in the projection factor decomposition ($ p = p_\mathrm{0} f_\mathrm{grad} f_\mathrm{o-g}$), the third component, $f_\mathrm{o-g}$, was extrapolated for the $\delta$~Scuti stars (not for $\alpha$~Lup, for which we used a dedicated hydrodynamical model), while the $f_\mathrm{grad}$ quantity was derived directly and independently from observations. We also recall that the geometric projection factor, $p_\mathrm{0}$, is directly linked to the limb-darkening and thus to the fundamental parameters of the stars. Second, for the Cepheids, the projection factors are consistent with the cross-correlation method  of the radial velocity determination. For DX~Cet and $\rho$~Pup, the line depth considered for the  $f_\mathrm{grad}$ correction is typical (i.e., averaged over all lines considered in our sample), which means that the derived projection factors are in principle very close to those to be used in  the cross-correlation method. For $\alpha$~Lup, AI~Vel, and $\beta$~Cas, there is no velocity gradient in the atmosphere and the projection factor is the same regardless of the line considered, and thus is consistent with the cross-correlation method. There remains only the fact that in the cross-correlation method a Gaussian is fitted to the cross-correlated spectral line profile, which in principle reduces the value of the projection factor by few percents compared with that based on the  first-moment method \citep{nardetto06}. Reducing the projection factor for short pulsators in Fig.~\ref{Pp_pulsating} by a few percents will basically not change the relative position of $\alpha$~Lup and our qualitative conclusion that the period-projection factor relation seems to straightly link $\delta$~Scuti stars and classical Cepheids. Third, the fast rotation of some $\delta$~Scuti stars (as well as the inclination of their rotation axis) can affect the projection factor by 10\%  or even more \citep{guiglion13}. However, the $\delta$ Scuti stars in our sample show $v_{eq} \sin i$ values lower than 13 \kms, except for $ \beta$~Cas  with $v_{eq} \sin i \simeq 75$ \kms.

\begin{figure}
   \centering
   \includegraphics[width=\columnwidth,height=\columnwidth]{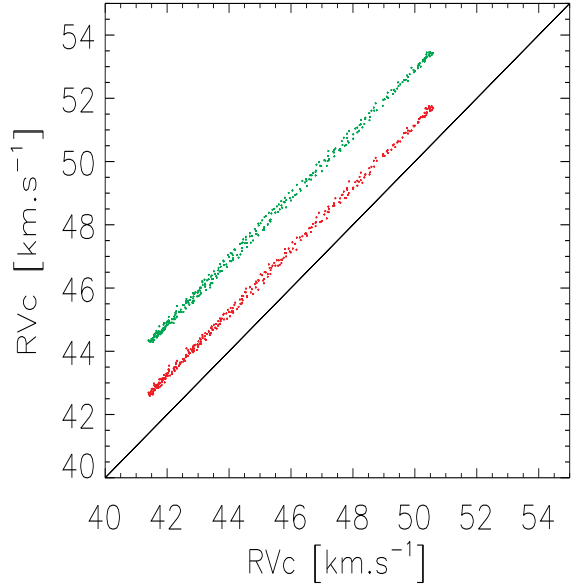}
   \caption{Van Hoof diagram for $\rho$~Pup. The radial velocity curve associated to the \ion{FeI} 5410.91 (green) and
\ion{FeI} 6393.60~\AA\ (red) spectral line profiles is plotted as a function of the same quantity, but for \ion{CrII} 4634.07. The solid line corresponds to the identity function. There is no Van Hoof effect detected (no loops), while the $\gamma$-velocity effect is clearly seen (y-axis offsets of the curves).}
              \label{f10}
    \end{figure}

Even if we were able to derive both the Baade-Wesselink projection factors and the centre-of-mass velocities for the eleven pulsating stars in our sample, an effort (both observational and theoretical) has still to be made to understand the spectral line asymmetry properly. For classical Cepheids, the hydrodynamical models reproduce the observed velocities and even the atmospheric velocity gradient very well \citep{fokin96, nardetto04, nardetto07}. Nevertheless, currently, there is no hydrodynamical model of pulsating stars that can correctly reproduce the $\gamma$-velocities and $\gamma$-asymmetries. In particular, we emphasize that the k-term quantity cannot be explained by a Van Hoof effect \citep{mathias93, mathias95}. In Fig.~\ref{f10}, we show the Van Hoof diagram for the three spectral lines of Sect. \ref{rpup} for $\rho$~Pup. 

The Van Hoof effect is not detected because there is no phase shift between the radial velocities of the different spectral lines (i.e., no loops). We also find that the two curves have a slope close to one, indicating a null atmospheric velocity gradient, as already mentioned. Nevertheless, the zero-points of the two curves are shifted, as a result from the $\gamma$-velocity offsets (Fig. \ref{ref_fig_rpup}, panel a). The
possible physical causes of these offsets (e.g., additional modes) are under analysis.

%%%%%%%%%%
\section{Conclusions}\label{s5}
%%%%%%%%%%

%\ep{ We presented HARPS high-resolution spectra of \rpup\, and \dxcet, which increases our sample of pulsating stars observed with HARPS (and for which we could measure the spectral lines asymmetry properly) to eleven (including eight Cepheids and one $\beta$~Cephei star, see Sect. \ref{s_new} for the detail).}

We presented HARPS high-resolution spectra of \rpup\, and \dxcet, thus  increasing our sample to eleven stars.
% of pulsating stars observed with HARPS to eleven.}
Among them, the $\beta$ Cep star $\alpha$~Lup shows an interesting  positive value of $2.2$ \kms\, for the k-term, while other kinds of pulsating stars have negative values (ranging from $-1$ to 0 \kms). We also found that $\rho$~Pup shows the same 2K velocity amplitude ($8.84\pm0.19$ \kms) regardless of  the spectral line considered in our sample (45 lines), while this is not the case for stars with amplitudes larger than  22.5~\kms\,, where a trend is indeed found as a function of the the line depth. This behaviour, if confirmed, would mean that the atmospheric velocity gradient correction on the Baade-Wesselink projection factor, $f_\mathrm{grad}$, is equal to 1.0 for any type of pulsating star as soon as its 2K velocity amplitude is lower than 22.5~\kms. Finally, we found that the period-projection factor relation might be common to classical Cepheids and $\delta$~Scuti, while the $\beta$ Cep star $\alpha$~Lup stands at 8$\sigma$ above the relation. 

The excitation of different pulsation modes (fundamental and overtone radial, pressure and gravity nonradial) 
usually makes the use of $\delta$~Sct stars as distance indicators problematic. For \rpup, 
our analysis of light and radial velocity curves identified the dominant mode as  the fundamental radial one. 
Because it is a very bright star with an accurate  {\it Hipparcos} parallax, \rpup\, can be now considered as the  best standard candle for the $\delta$ Sct class.
The {\it Hipparcos} absolute magnitude gives us a luminosity $\log L/L_{\sun}$=1.38. Combined with
 $T_{\rm eff}$=6650-6800~K, this implies that \rpup\, is very close to the red border of the instability strip,
probably leaving the central part where most of the HADS stars are located \citep{pamya}.

Moreover, from the  {\it Hipparcos} parallax ($\pi = 7.02\pm0.17$ mas), the mean radius of $R=3.52\pm0.07\Ro$ \citep{antoci13}, and our value of the projection factor $p=1.36$ together with our radial velocity curve (in particular, $2K=8.8$~\kms), we were able to infer  an absolute angular diameter  variation of 11~$\mu$as (or 0.7\%).  Unfortunately, there are no K photometric observations of \rpup\, 
that cover a full pulsation cycle to which one might apply the infrared surface brightness relation and confirm these findings.

\begin{acknowledgements}
The authors wish to thank the anonymous referee for useful comments.
EP and MS acknowledge financial support from the Italian PRIN-INAF 2010
{\it Asteroseismology: looking inside the stars with space- and ground-based
observations}. MR scknowledges financial support from the FP7 project 
{\it SPACEINN: Exploitation of Space Data for Innovative Helio-
and Asteroseismology}. VSS is an Aspirant PhD fellow of the Fonds voor Wetenschappelijk Onderzoek, Vlaanderen (FWO), Belgium. NN  and EP acknowledge the {\it Observatoire de la C\^ote d'Azur}
for the one-month grant that has allowed EP to work at OCA in April and May 2013.
\end{acknowledgements}

%-------------------------------------------------------------------

\end{document}